\documentclass[a4paper,12pt]{article}

\usepackage{amssymb}
\usepackage{amsmath}
\usepackage{graphicx}
\usepackage{bm}
\usepackage[english]{babel}

\newcommand{\rmd}{{\rm d}}
\newcommand{\rmi}{{\rm i}}
\newcommand{\rme}{{\rm e}}
\newcommand{\Or}{{\rm O}}
\newcommand{\eref}[1]{(\ref{#1})}

\renewcommand{\vec}[1]{\bm{#1}}

\begin{document}

\title{Whole-plane self-avoiding walks and\\ radial Schramm-Loewner evolution:\\ a numerical study}

\author{\\Marco Gherardi
\\{\small\it Dipartimento di Fisica and INFN --- Sezione di Milano I}
 \\[-0.2cm]{\small\it Universit\`a degli Studi di Milano}
 \\[-0.2cm]  {\small\it Via Celoria 16, I-20133 Milano, Italy}
 \\[-0.2cm]  {\small e-mail: {\tt Marco.Gherardi@mi.infn.it}}
 \\
}

\maketitle

\begin{abstract}
We numerically test the correspondence between the scaling limit of self-avoiding walks (SAW) in the plane and Schramm-Loewner evolution (SLE) with $\kappa=8/3$. We introduce a discrete-time process approximating SLE in the exterior of the unit disc and compare the distribution functions for an internal point in the SAW and a point at a fixed fractal variation on the SLE, finding good agreement. This provides numerical evidence in favor of a conjecture by Lawler, Schramm and Werner. The algorithm turns out to be an efficient way of computing the position of an internal point in the SAW.
\end{abstract}

\section{Introduction}

Schramm-Loewner evolution (SLE) is a family of random processes on conformal maps, which gives rise to a one-parameter family of measures on curves satisfying conformal invariance.
It was first introduced in \cite{Schramm} to describe the scaling limit of loop-erased random walks, but it was soon found to correspond to a large family of geometrical objects defined in the context of lattice models.
The latter include interfaces in critical models --- e.g. the interface between phases in the Ising model or the boundary of the percolating cluster in critical percolation --- and walk models, such as the self-avoiding walk in the scaling limit.
A wealth of results have been obtained for SLE in the \emph{chordal} geometry, which involves curves starting and ending on points lying on the boundary of some connected domain, and in the \emph{radial} geometry, where one of the points is on the boundary and the other is in the bulk.
Much less has been proved or checked in the whole-plane geometry, where both the starting and ending points lie in the bulk, such as the points $0$ and $\infty$ in $\mathbb C$.

In this paper we consider SLE in the whole plane and compare it with the planar self-avoiding walks (SAW).
The correspondence between the two models has been conjectured by Lawler, Schramm and Werner in \cite{LSW:planarSAW}, on the basis of restriction covariance and conformal invariance.
In particular, we will compare the scaling forms of the distribution functions in the two models.
For this purpose, one has to be careful about which points along the curves have to be considered.
On the SLE side, a suitable parametrization must be chosen; we will rely on parametrization by \emph{fractal variation}, and will pick the point at a fixed value of this parametrization.
On the SAW side, we are going to consider an internal point deep inside the walk, so as to avoid finite-chain effects.
Thanks to the scaling relation between the fractal variation and the natural parametrization of SLE, and to the fact that the discrete chains are sampled independently, this turns out to be a potentially efficient method for computing the distribution function of an internal point in the SAW, since it produces independent samples of length $L$ in time $\Or(L^{1.5})$ with room for easy improvement.

The plan of the paper is as follows.
In section \ref{sec:definitions} we quickly review the relevant facts and definitions about Schramm-Loewner evolutions and self-avoiding walks; in section \ref{sec:discrete} we introduce and discuss the discrete process we will be simulating in the whole plane; in section \ref{sec:distribution} we define the distribution function and its conjectured scaling behavior; in section \ref{sec:results} we present our numerical results; the appendix deals with the scaling of the average step length and the hull size in discrete SLE.

\section{\label{sec:definitions}Definitions and background}

SLE is a widely studied and reviewed subject, see e.g. \cite{Cardy,KagerNienhuis} for an introduction.
Here we will only recall the main ingredients, in order to make this paper as self-contained as possible.

SLE is a stochastic differential equation describing the evolution of a parametrized family of conformal maps $g_t$ in a domain $\mathcal D$.
The actual form of the equation depends on the geometry, but in general it is written in terms of a function taking values on the boundary of $\mathcal D$, called \emph{driving function}.
The driving function is rescaled Brownian motion $\sqrt{\kappa}B_t$ living on the boundary; $\kappa$ is a positive constant.
The maps $g_t$ map some ($t$-dependent) subset of $\mathcal D$ onto $\mathcal D$, so that one can define the growing \emph{hull} $K_t$ as the domain of $g_t$, i.e. the set of points for which the differential equation still has a solution up to time $t$.
The hull turns out to be \emph{generated} by a curve $\gamma(t)$, i.e. it is the union of the image of the curve with the interior of any loop it has closed.
The curve --- also called the \emph{trace} of the process --- is the pre-image of the driving function under $g_t$:
\begin{equation}
g_t\left(\gamma(t)\right)=\sqrt{\kappa}B_t.
\end{equation}
As the variance $\kappa$ of the Brownian driving function varies, the properties of $\gamma(t)$ change dramatically, and very different models have been proved or conjectured to be described by SLE for some value of $\kappa$ in the scaling limit. 
SLE has proved a useful tool for addressing very diverse problems and questions (see for instance \cite{Gruzberg, Duplantier} and references therein).

Interesting issues arise when one considers the reparametrization of $\gamma(t)$ \cite{Lawler:reparam}.
The SLE equation generates the hulls with a natural parametrization, which corresponds to a linearly growing \emph{capacity} (which is a property related to the expansion of $g_t$ around some special point and is a measure of the ``conformal size'' of the hull, see \cite{Landkof} for a precise definition).
As long as one is interested in parametrization-independent observables this is not an issue. 
But we will be considering the distribution function of the point $\gamma(t)$ for some fixed $t$, so the choice of parametrization is clearly essential here.
One way of dealing with this problem in the context of a numerical work is to generate the curves already with the correct parametrization --- here \emph{correct} means the one corresponding to the parametrization \emph{by length} of the SAW.
The technical problem of producing discrete SLE traces parametrized by length will be treated elsewhere.
In this paper instead we will take advantage of a technique introduced by Kennedy \cite{Kennedy:lengthofanSLE}, which consists in defining a suitable notion of length along the discretized curve, and using it as the parametrization (see section \ref{sec:discrete}).

A self-avoiding walk $\omega$ on the square lattice with a fixed number of steps $N$ is an ordered collection $\omega=\left\{\vec{\omega}_0,\ldots,\vec{\omega}_N\right\}$ with $\vec{\omega}_i\in\mathbb Z^2$, such that $\left|\vec{\omega}_i-\vec{\omega}_{i-1}\right|=1$ and $\vec{\omega}_i\neq\vec{\omega}_j$ for $i\neq j$.
The SAW model is the uniform measure on these objects, i.e. on all non-intersecting $N$-step nearest-neighbor walks \cite{MadrasSlade}.

A numerical check of the equivalence between the distribution functions of SLE and SAW in the \emph{half plane} has been carried out in \cite{Kennedy:lengthofanSLE}, with positive results.

\section{\label{sec:discrete}Discrete whole-plane SLE}

Simulation of SLE in the half plane $\mathbb H$ is usually based on the discrete process introduced in \cite{Bauer:DSLE}, where it is also proved that it converges weakly to SLE.
Here we introduce a whole-plane version of this discrete process.

The idea (see also \cite{Kennedy:fastalgorithm}) is to write the Loewner map as a composition of (finitely many) conformal maps, chosen from a simple parametrized family.
Time is partitioned by letting $0=t_0<t_1<\ldots$.
The driving function is defined in such a way as to be equal to $\sqrt{\kappa}B_{t_k}$ at the special times $t_k$ and constant in between.
This amounts to approximating the Brownian motion with a piecewise constant function, so that the incremental maps $G_k=g_{t_k}\circ g_{t_{k-1}}^{-1}$ are obtained by solving the Loewner equation with constant driving function up to time $t_k-t_{k-1}$.

Let us denote $\xi_t\equiv\sqrt{\kappa}B_t$. 
Consider the radial SLE equation
\begin{equation}
\label{eq:radialSLE}
\frac{\rmd}{\rmd t}\tilde g_t(z)=\tilde g_t(z)\frac{\exp(\rmi\xi_t)+\tilde g_t(z)}{\exp(\rmi\xi_t)-\tilde g_t(z)},\qquad \tilde g_0(z)=z
\end{equation}
which describes a hull starting from $\exp{(\rmi \xi_0)}$ and growing towards the origin inside the unit disc $\mathbb D$.
In analogy with the half-plane case, the hull $K_t$ at time $t$ is defined as the set of those points in $\mathbb D$ for which the differential equation does not admit a solution that exists up to time $t$.
By composition with the complex inversion $(z\mapsto 1/z)$ one can define the map 
\begin{equation}
\label{eq:reciprocalmap}
g_t(z)=\frac{1}{\tilde g_t\left(1/z\right)}
\end{equation}
which now describes a hull growing in $\mathbb C\setminus\mathbb D$ from $\exp{\left(-\rmi\xi_0\right)}$ to $\infty$.
The new map in \eref{eq:reciprocalmap} happens to satisfy the same equation \eref{eq:radialSLE}, again with initial condition $g_0(z)=z$, but now with driving function $-\xi_t$, which has the same law as $\xi_t$ if started from $\xi_0=0$.
This process is called \emph{radial SLE growing to infinity}.
Notice that $g_t$ does not describe a curve truly in the \emph{whole plane}, since the unit disc is a forbidden region.
However, we are going to focus on the large-scale regime in the following, and expect the cutoff at length scale $1$ to be irrelevant in this limit.

The discrete process is based on the discretization of the evolution of the \emph{inverse} map $g_t^{-1}$.
This map \emph{grows} the hull at time $t$, i.e. it maps $\mathbb C\setminus\mathbb D$ onto the complement of the hull in $\mathbb C\setminus\mathbb D$.
The atomic step of the discretization will be performed by a map --- which we are going to call \emph{incremental} --- that grows a \emph{slit} (a small radial segment) out of the unit disc.
Obtaining this map is simply a matter of solving \eref{eq:radialSLE} in the special case when the driving function is constant, thus finding the Loewner map corresponding to a straight line growing towards infinity; the incremental map is its inverse.
The result is\footnote{
Some care must be taken in choosing the right sign before the square root and the position of the branch cut of the square root itself, so that a point outside $\mathbb D$ be mapped to a point outside $\mathbb D$.
The conformal map in \eref{eq:incrementalmap} already appeared in the literature about diffusion limited aggregates \cite{HastingsLevitov}, where it was used to represent the attachment of a single grain onto the cluster.}
\begin{equation}
\label{eq:incrementalmap}
\phi_t(z)=\frac{1}{2\rme^{-t}z}\left[(z+1)^2-2\rme^{-t}z-(z+1)\left((z+1)^2-4\rme^{-t}z\right)^{1/2}\right].
\end{equation}
The greater the time $t$ is, the longer will be the line grown by $\phi_t$.
In the language of potential theory, one says that the hull grown by $\phi_t$ has \emph{logarithmic capacity} $t$.

Brownian motion driving the evolution lives on the boundary of the standard domain, which is the upper half plane in the chordal case, the unit disc in the radial case and the complement of $\mathbb D$ in $\mathbb C$ in the case at hand.
Notice that, by translation invariance, chordal SLE in $\mathbb H$ growing from $\xi_0$ to $\infty$ is chordal SLE growing from $0$ to $\infty$ translated by $\xi_0$, so that in simulating the discrete process one usually translates $\xi_t$ back to the origin after each iteration.
Translations are replaced by rotations in the radial case.
Therefore, at each step we will want to rotate $\exp{(\rmi \xi_t)}$ back to $1$.
The whole discretized hull will be generated by alternately composing an incremental map with a rotation.
Let us call $\Delta_k$ the logarithmic capacity of the incremental map at the $k$-th step (i.e. the time at which the $k$-th map is evaluated), and correspondingly let $\delta_k$ denote the angle of the $k$-th rotation.
The incremental map and the rotation themselves will then be denoted $\phi_{\Delta_k}$ and $R_{\delta_k}$ respectively, where $R_{\delta_k}(z)=z\exp{\left(\rmi\delta_k\right)}$.
Let $\gamma_k$ be the image of $1$ under the composed map at step $k$
\begin{equation}
\label{eq:composition}
\gamma_k=R_{\delta_1}\circ\phi_{\Delta_1}\circ\cdots\circ R_{\delta_k}\circ\phi_{\Delta_k}(1).
\end{equation}
Notice that the order in which the maps are composed is the opposite as the usual one; this is essentially a consequence of the fact that we are discretizing the inverse Loewner map.
The approximate curve we are interested in is embodied by the collection of points $\left\{\gamma_k\right\}$.

\begin{figure}[t]
\centering
\includegraphics{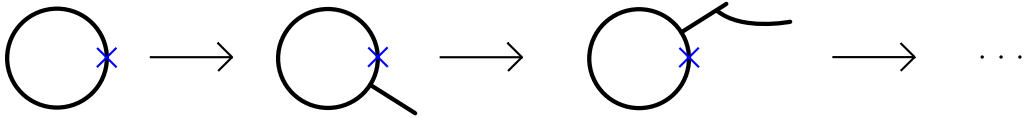}
\caption{(Color online) A visualization of the discrete process. The leftmost arrow corresponds to the action of $R_{\delta_k}\circ\phi_{\Delta_k}$, the second corresponds to $R_{\delta_{k-1}}\circ\phi_{\Delta_{k-1}}$, and so on. The blue cross is the point $1$, where the next slit will be based.}
\label{fig:composition}
\end{figure}

Refer to figure \ref{fig:composition}.
The first map grows a slit based at point $1$, which is then rotated to $\exp{(\rmi\delta_k)}$.
The second map again grows a slit based at $1$, so that the old slit will be sent away from the unit disc, and will be based somewhere between $1$ and the tip of the new slit (notice that the old slit will not in general retain its rectilinear shape).
Then the whole hull is rotated by $\delta_{k-1}$ and another slit is grown.
When the last composition with $R_{\delta_1}\circ\phi_{\Delta_1}$ is reached, the $k$-th point on the trace is found.
The process is repeated for each point $\gamma_k$ that has to be computed.

The resulting trace depends on the choice of the time-like parameters $\left\{\Delta_k\right\}$ and the space-like parameters $\left\{\delta_k\right\}$.
In analogy with the chordal half-plane case \cite{Bauer:DSLE} we choose to draw each $\delta_k$ as a Bernoulli variable in the set $\left\{-\sqrt{\kappa\Delta_k},+\sqrt{\kappa\Delta_k}\right\}$.
This choice amounts to approximating the Brownian motion with a piecewise constant function, the relation between $\Delta_k$ and $\delta_k$ reproducing the well-known space-time scaling of Brownian motion with variance $\kappa$.
One has additional freedom in choosing the time intervals $\left\{\Delta_k\right\}$.
Changing the latter corresponds to a reparametrization of the resulting trace.
The uniform partition $\Delta_k=\Delta$ yields the parametrization \emph{by capacity}, i.e. the natural SLE parametrization where the hull increases its capacity linearly in time.
The choice of parametrization is crucial for our purposes, since we are interested in the spatial distribution function, which is a parametrization-dependent observable.
In order to reproduce the correct parametrization (that corresponding to the natural parametrization of the supposed scaling limit of self-avoiding walks) we use the following method.
First, we fix a scale $\lambda$.
Then, at each step, we compute the \emph{fractal variation} of the curve as follows.
Let $k_0=0$.
Times $\left\{k_i\right\}_{i=1\ldots n}$ are defined recursively: given $k_i$, $k_{i+1}$ is the first time after $k_i$ such that $\left|\gamma_{k_{i+1}}-\gamma_{k_i}\right|\geq\lambda$.
The fractal variation at step $k$ is defined as
\begin{equation}
\label{eq:fractalvariation}
{\rm var}_\lambda\left(\left\{\gamma_0,\ldots,\gamma_k\right\}\right)=n\lambda^{d_f}.
\end{equation}
where $n$ is the largest integer such that $k_n\leq k$, and $d_f$ is the fractal dimension of continuum SLE as a function of $\kappa$ \cite{Beffara}
\begin{equation}
\label{eq:fractaldimension}
d_f=1+\frac{\kappa}{8}.
\end{equation}
The growth process will be stopped when ${\rm var}_\lambda$ reaches a fixed value $\Upsilon$.
This procedure has the advantage of being independent of the original parametrization of $\gamma$.
Nonetheless, it is sensitive to discretization problems, for instance when the steps taken by the trace $\gamma_k$ become too wide with respect to $\lambda$.
A uniform partition $\Delta_k=\Delta$ causes the average step length at step $k$, $l_k=\left<\left|\gamma_k-\gamma_{k-1}\right|\right>$, to diverge\footnote{
This is true in the whole plane.
The opposite happens in the half plane, where the average step length converges to zero.}
as the number of steps $k$ grows (the average is over all realizations of the process up to step $k$).
To avoid approximation problems we choose the time intervals $\Delta_k$ in a non-uniform fashion, having them scale as
\begin{equation}
\label{eq:Deltascaling}
\Delta_k\sim k^{-1}
\end{equation}
so as to compensate for the divergence of the average step length (see the appendix for details).
We can then define $l=l_k$, at least for $k$ large enough.

The discrete process is defined operatively as follows:
\begin{enumerate}
\item Fix the three scales $l$, $\lambda$ and $\Upsilon$; start with $k=1$
\item\label{item1} Choose $\Delta_k$ according to \eref{eq:Deltascaling}
\item Draw $\delta_k$ uniformly in $\left\{-\sqrt{\kappa\Delta_k},+\sqrt{\kappa\Delta_k}\right\}$
\item Calculate $\gamma_k$ as in \eref{eq:composition}
\item Measure the fractal variation \eref{eq:fractalvariation}
\item If ${\rm var}_\lambda\left(\left\{\gamma_0,\ldots,\gamma_k\right\}\right)\geq\Upsilon$ stop, otherwise increase $k$ and return to \ref{item1}.
\end{enumerate}

Notice that the time when the fractal variation reaches $\Upsilon$ depends on the scale $\lambda$ it is measured at, which should be sent to 0 in order to obtain the true fractal variation.
Of course, the curve obtained by means of the discrete process defined above is not really fractal at all, but it displays fractal properties only at a large enough scale.
Therefore, values of $\lambda$ much less than the average step length are expected to yield the trivial parametrization.
On the other hand when $\lambda$ becomes comparable to $\Upsilon^{1/d_f}$, ${\rm var}_\lambda$ suffers from rounding problems.
We will keep $\lambda$ between these two cutoffs, and study how results depend on this choice.

\begin{figure}[t]
\centering
\includegraphics[scale=0.8]{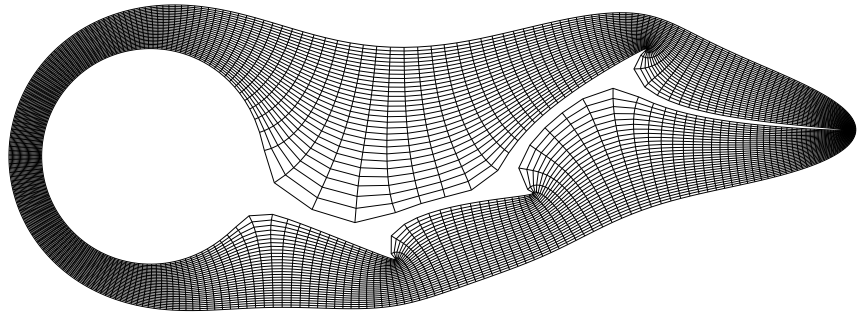}
\caption{A visualization of the composed map in \eref{eq:composition} after $4$ iterations. The curves are the images of circles and radii outside the unit disc.}
\label{fig:map}
\end{figure}

An example of the composed map after a few iteration is in figure \ref{fig:map}.

\section{\label{sec:distribution}Distribution function}

Let us consider the $M$-th point in an $N$-step SAW $\omega$.
The probability distribution $P_{N,M}(\vec{r})$ of this internal point (the probability that it lies at site $\vec{r}$ on the lattice) has the following scaling form in the limit $N,M\to\infty$, $\vec{r}\to\infty$, with $M/N$ fixed:
\begin{equation}
P_{N,M}(\vec{r})\sim\frac{1}{\xi_M^2}f_{\rm SAW}\left(\rho,\frac{M}{N}\right)
\end{equation}
where $\rho=\left|\vec{r}/\xi_M\right|$ and $\xi_M^2=\left<\left|\vec{\omega}_M-\vec{\omega}_0\right|^2\right>$ is the \emph{correlation length} after $M$ steps.
The average $\left<\cdot\right>$ is on the self-avoiding walk ensemble.
The universal function $f_{\rm SAW}$ is the \emph{renormalized distribution function} we are going to compare with its SLE analogue.
It depends on the ratio $M/N$ --- i.e. on how much the $M$-th point feels the finiteness of the chain.
Since we want the distribution of a point a finite distance away from the origin in a truly infinite curve, we should take $M/N\to 0$.

Let us then define the corresponding quantities for the SLE approximated trace $\left\{\gamma_k\right\}$ $(k=0,\ldots,N_{\Upsilon})$, where $N_{\Upsilon}$ is the number of points computed on the trace up to the stopping time when the fractal variation has reached $\Upsilon$.
We are not considering the full distribution $P(z)$, since the discrete process as defined above explicitly breaks rotational invariance --- it has $1$ as a special point.
Instead, consider the probability density $P_{\Upsilon}(r)$ that the point $\gamma_k$ has modulus $r$ when $k=N_{\Upsilon}$, then we will suppose it has the following scaling behavior in the \emph{long-chain limit}, i.e. when $\Upsilon\to\infty$ and $r\to\infty$ with step length $l$ fixed:
\begin{equation}
\label{eq:SLEdistributionfunction}
P_{\Upsilon}(r) \sim \frac{1}{\xi^2_{\Upsilon}}\;f_{\rm SLE}(\rho)
\end{equation}
where $\rho=r/\xi_{\Upsilon}$, and $\xi_{\Upsilon}$ is the correlation length
\begin{equation}
\xi^2_{\Upsilon}=\left< \left| \gamma_{N_\Upsilon} \right|^2 \right>
\end{equation}
(the average $\left<\cdot\right>$ here is over all realizations of the driving function).
Implicit in this conjectured behavior is the assumption that the presence of the unit disc as a forbidden region be irrelevant in the long-chain limit defined above.
Scaling form \eref{eq:SLEdistributionfunction} for the distribution function will be verified \emph{a posteriori}.
Notice that of course $f_{\rm SLE}$ implicitly depends on $\kappa$.

In order to quantitatively compare the distributions we will focus on their moments.
It is convenient to introduce an infrared cutoff $\rho_{\rm MAX}$ --- i.e. a window for computing the moments --- since deviations from scaling are more pronounced in the large-$\rho$ regime.
Suppose we have sampled $\mathcal N$ instances $z_i$ $(i=0,\ldots,\mathcal N-1)$ of one point on the chain --- be it an internal point inside a SAW or the point at a fixed fractal variation on the approximated SLE trace.
We will compute the following quantities:
\begin{equation}
\label{eq:moments}
M_{2k}=\frac{\sum' \left|z_i\right|^{2k}}{\left[\sum'\left|z_i\right|^2\right]^k}
\end{equation}
where the sums $\sum'$ are restricted to the window
\begin{equation}
\frac{\left|z_i\right|}{\xi_\Upsilon}<\rho_{\rm MAX}.
\end{equation}
We will choose $\rho_{\rm MAX}=3$ in the following, which only leaves out the tail of the distribution (see figure \ref{fig:distributionfunctions}).

\section{\label{sec:results}Numerical results}

We simulated an ensemble of $10^6$ self-avoiding walks of length $N=100\,000$ using the pivot algorithm \cite{MadrasSokal,Kennedy:Pivot}.
We considered an internal point with $M=8000$
(detailed analysis of the systematic error due to the finiteness of $M/N=0.08$ shows that it is negligible when compared to the deviations in discrete-SLE data due to the finiteness of $\lambda$ and $\Upsilon$).
The discrete SLE process was simulated for several values of $\lambda$ and $\Upsilon$, generating $\sim 100\,000$ independent samples for each choice.
In the following we are going to fix the average step size $l$ and measure everything else in units of $l$.

The time needed to compute the $k$-th point along the trace through \eref{eq:composition} is proportional to $k$.
Therefore, generating a $N$-step chain requires a time of order $N^2$, which can be interpreted by saying that the time-per-point is $\Or(N)$.
As shown in the appendix, partitioning time as in \eref{eq:Deltascaling} causes the fractal variation to scale as $\Upsilon\sim N^{d_f}$.
All together, the algorithm described here generates chains of length $\Upsilon$ in time $\Or(\Upsilon^{2/d_f})$, which is $\Or(\Upsilon^{3/2})$ for $\kappa=\frac{8}{3}$.
The time-per-point can be probably further improved by approximating the incremental map by its truncated Laurent series (see \cite{Kennedy:fastalgorithm}, where the numerical analysis is carried out in the half-plane geometry, suggesting that the time-per-point is $\Or(N^{0.4})$).
We shall not do that here, but this suggests that this algorithm might get close to generating independent samples of length $\Upsilon$ in time $\Or\left(\Upsilon\right)$, and even faster for higher values of $\kappa$.
Producing $100\,000$ samples with $\Upsilon=400$ took about 1000 hours on an Intel Pentium 4 with 1.80 GHz CPU\footnote{
This is the performance of a non-optimized code.}.

\begin{figure}[t]
\centering
\includegraphics{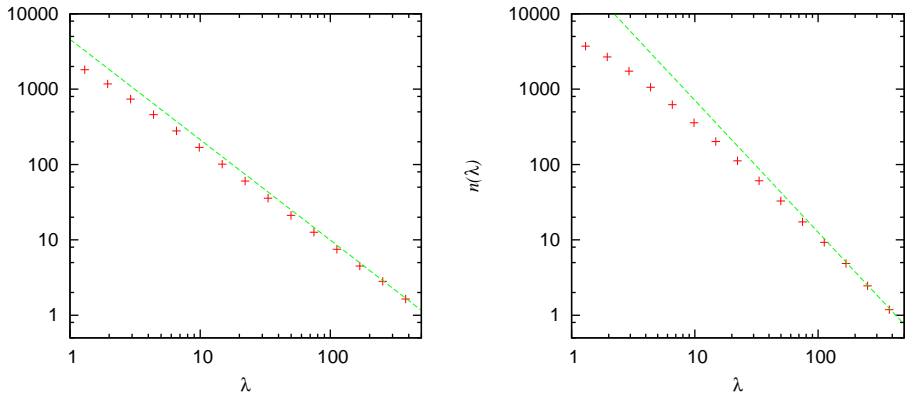}
\caption{(Color online) The number of segments of length $\lambda$ needed to cover the trace (red crosses), for $\kappa=\frac{8}{3}$ (left) and $\kappa=6$ (right). Dashed green lines represent the theoretical slope corresponding to fractal dimension $1+\kappa/8$. (Error bars are smaller than the crosses representing data).}
\label{fig:fractaldimension}
\end{figure}

As a preliminary test, we checked that the discrete SLE approximants approached the expected fractal dimension \eref{eq:fractaldimension}, since knowing its precise value is crucial when computing the fractal variation\footnote{
Notice that the fractal dimension does not depend on the parametrization.
Actually, we reverted to parametrization by capacity when testing $d_f$ --- i.e. we stopped the discrete process at a fixed number of iterations, without computing the variation.}
as in \eref{eq:fractalvariation}.
For increasing values of $\lambda$ we measured the number $n(\lambda)$ of segments of length $\lambda$ that are needed to cover up the entire trace, in the same fashion as when computing the fractal variation.
The expected behavior is $n(\lambda)\sim\lambda^{-d_f}$.
We checked this for several values of $\kappa$, finding good agreement.
Figure \ref{fig:fractaldimension} shows the results for two different values of $\kappa$.
For small $\lambda$ one sees the crossover to the true fractal dimension of the discrete trace, which is of course $1$.

\begin{figure}[t]
\centering
\includegraphics{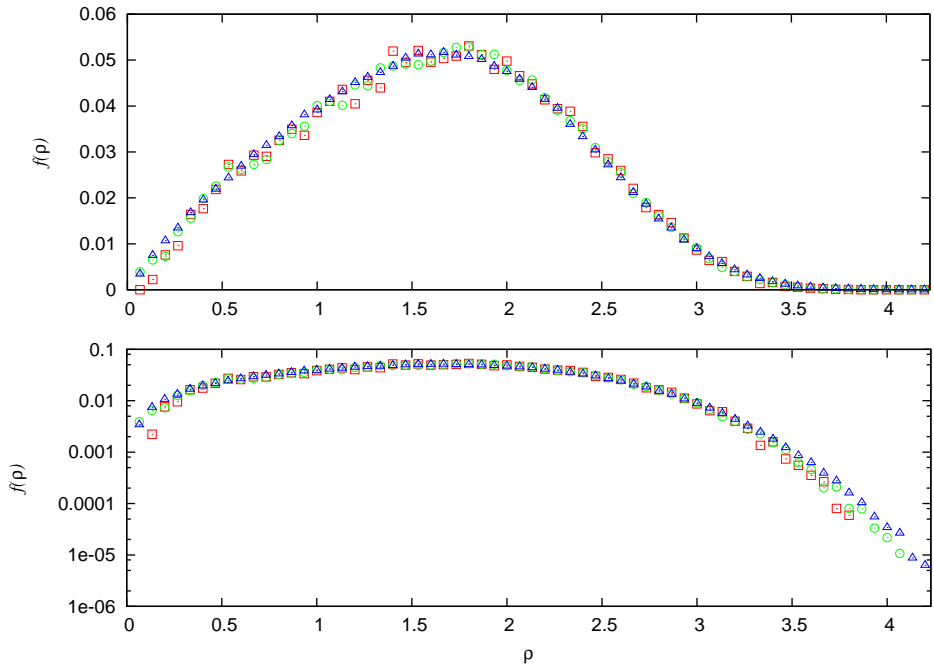}
\caption{(Color online) The renormalized distribution functions of SLE with $\Upsilon=400$ (red squares) and $\Upsilon=800$ (green circles), and that of a point inside a SAW (blue triangles).
The two plots display the same data; the one below has logarithmic scale on the y-axis.}
\label{fig:distributionfunctions}
\end{figure}

We are now ready to check whether $f_{\rm SAW}\equiv f_{\rm SLE}$ for $\kappa=\frac{8}{3}$.
Figure \ref{fig:distributionfunctions} is a comparative plot of the renormalized distribution functions for the SAW and the discrete whole-plane SLE, for two values of the variation, $\Upsilon=400,800$ (an averaging procedure has been adopted here, in order to smoothen oscillations due to the lattice).
The distributions all fall onto the same universal curve, apart from corrections to scaling in the extreme regimes.
To quantitatively check that this is the case, the moments \eref{eq:moments} have been computed for each value of $\Upsilon$ and for several values of $\lambda$; they are reported in table~\ref{tab:moments}, together with their SAW values.
\begin{table}[tp]
\centering
\begin{tabular}{c@{\extracolsep{0.2cm}}c@{}c@{}c@{}c@{}c@{}c}
\noalign{\hrule height 1pt}
& \multicolumn{4}{c}{$\Upsilon=400$} & $\Upsilon=800$ & \\
\cline{2-5}
& $\lambda=4$ & $\lambda=10$ & $\lambda=20$ & $\lambda=40$ & $\lambda=20$ & SAW\\
\hline
$M_4$ & 1.304(10) & 1.315(7) & 1.320(11) & 1.318(7) & 1.322(9) & 1.330(2)\\
$M_6$ & 1.970(18) & 2.009(14) & 2.022(25) & 2.008(14) & 2.030(20) & 2.059(5)\\
$M_8$ & 3.267(38) & 3.374(32) & 3.404(55) & 3.352(31) & 3.428(42) & 3.508(10)\\
$M_{10}$ & 5.787(84) & 6.045(70) & 6.11(12) & 5.955(68) & 6.177(93) & 6.379(23)\\
$M_{12}$ & 10.74(18) & 11.35(16) & 11.48(27) & 11.07(15) & 11.66(21) & 12.155(52)\\
\hline
$M_{12}$ & 15.77(32) & 16.94(27) & 16.79(46) & 14.92(23) & 17.43(37) & 18.763(94)\\
\noalign{\hrule height 1pt}
\end{tabular}
\caption{The first five non-trivial moments of the distribution function for whole-plane SLE and SAW. $\Upsilon$ is the fractal variation computed at scale $\lambda$. (Here $\rho_{\rm MAX}=3$, except for the bottom line where $\rho_{\rm MAX}=3.5$).}
\label{tab:moments}
\end{table}
Accordance for low-order moments is more easily established, while inspection of high-order ones helps in recognizing systematic deviations.
Let us fix $\Upsilon=400$ first.
Values for $\lambda=40$ suffer from severe deviations due to the coarse grained nature of the procedure used to compute the fractal variation. 
On the other hand, a systematic drift in $\lambda$ is present for smaller values.
The best compromise seems to lie in the middle; we will then fix $\lambda=20$ and increase $\Upsilon$.
The moments for $\Upsilon=800$ all lie within two standard deviations from the SAW values, apart from the highest-order one which is close.
Deviations are apparent, the SAW values being systematically larger than the others, but again SLE moments keep increasing as $\Upsilon$ is doubled, thus approaching the expected values.
As expected, these systematic corrections get larger if the cutoff $\rho_{\rm MAX}$ is increased.
Convergence for $\rho=3.5$ (see $M_{12}$ in the bottommost line in table \ref{tab:moments}) is slower but still consistent.

\section{Conclusions}

We have introduced a discrete process approximating radial Schramm-Loew\-ner evolution growing to infinity, by the iteration of conformal maps defined outside the unit disc.
On one hand, we have considered the distribution function of a point on the trace at a fixed value of the fractal variation.
As the fractal variation reaches infinity the distribution forgets about the presence of the forbidden region $\mathbb D$.
On the other hand, we measured the distribution function of a point deep inside a whole-plane self-avoiding walk.
When $\kappa=\frac{8}{3}$, the two universal functions match, thus providing evidence that the scaling limit of SAW is whole-plane SLE.
Moreover, computing the position of an internal point in a SAW by exactly sampling the discrete SLE process seems to be an efficient algorithm, which is open to further improvement.

\section*{Acknowledgements}
The author wishes to thank Tom Kennedy for useful discussions and Sergio Caracciolo for helpful suggestions and a careful reading of the manuscript.

\appendix
\section*{Appendix}
\setcounter{section}{1}

We derive here the scaling form \eref{eq:Deltascaling}.
Let $l_k$ denote the length of the $k$-th step in the discretized growth, that is
\begin{equation}
\label{eq:app:ldef}
l_n=\left|\gamma_n-\gamma_{n-1}\right|.
\end{equation}
The exponential of the logarithmic capacity is a measure of the linear size of the growing hull. 
This is essentially a consequence of the Koebe 1/4 theorem, which bounds the size of the image of the unit disc $\mathbb D$ under a conformal map $g$ in terms of $g'(0)$.
Then, by approximating \eref{eq:app:ldef} by
\begin{equation}
l_n\sim\left|\gamma_n\right|-\left|\gamma_{n-1}\right|
\end{equation}
as if it grew radially, one has
\begin{equation}
\label{eq:app:l}
\begin{split}
l_n &\sim \exp\left(\sum_{k=1}^n\Delta_k\right) - \exp\left(\sum_{k=1}^{n-1}\Delta_k\right)\\
&= \left(e^{\Delta_n}-1\right)\exp\left(\sum_{k=1}^{n-1}\Delta_k\right).
\end{split}
\end{equation}
Setting
\begin{equation}
\label{eq:app:position}
\Delta_k=\frac{\Delta}{k}
\end{equation}
in \eref{eq:app:l} gives
\begin{equation}
\label{eq:app:l2}
l_n \sim \left(\exp\frac{\Delta}{n}-1\right) \exp\left(\Delta H_{n-1}\right)
\end{equation}
where the $H_n$'s are the harmonic numbers, whose expansion in $n$ is
\begin{equation}
\label{eq:app:harmonicnumbers}
H_n=\ln n + \gamma + \Or\left(\frac{1}{n}\right)
\end{equation}
($\gamma$ is the Euler-Mascheroni constant).
By substituting in \eref{eq:app:l2} and expanding both exponentials in $n$ one gets
\begin{equation}
l_n\sim \left[\frac{\Delta}{n}+\Or\left(\frac{1}{n^2}\right)\right]
\left(n-1\right)^\Delta \rme^{\Delta\gamma} \left[1+\Or\left(\frac{1}{n}\right)\right].
\end{equation}
By choosing $\Delta=1$ one finally obtains 
\begin{equation}
l_n \sim \rme^\gamma \left[ 1+\Or\left(\frac{1}{n}\right)\right]
\end{equation}
which shows that scaling $\Delta_k$ as in \eref{eq:Deltascaling} provides an asymptotically constant step length.
Contrary to the half-plane case --- where a similar computation shows that an approximately constant $l_n$ is obtained by choosing $\Delta_k=k\Delta$ --- there is no freedom left here to choose the average step length, since $\Delta$ has to be fixed to $1$ in order to have the correct scaling behavior.
Operatively, we keep a constant $\Delta_k$ for a few steps, until the step length has approximately reached the desired value $l$, and we set $\Delta_k=1/k$ from then on.

In order to obtain a scaling relation for the average number of steps $n$ needed to reach fractal variation $\Upsilon_n$, again we use the exponential of the logarithmic capacity as a measure of the chain size, which by \eref{eq:fractalvariation} is of order $\Upsilon_n^{1/d_f}$, so that for $n$ large we have
\begin{equation}
\exp \sum_{k=1}^n\Delta_k \sim \Upsilon_n^{\frac{1}{d_f}}
\end{equation}
which, by substituting $\Delta_k$ given by \eref{eq:app:position} with $\Delta=1$ and again using \eref{eq:app:harmonicnumbers}, yields
\begin{equation}
n \sim \Upsilon^{\frac{1}{d_f}}.
\end{equation}

\end{document}